\documentclass[aps,pra,showpacs,twocolumn,superscriptaddress]{revtex4}
\usepackage{amsmath}
\usepackage{amsfonts}
\usepackage{amssymb}
\usepackage{graphicx}
\usepackage{bbm}
\usepackage{natbib}
\usepackage[dvipdfm]{hyperref}
\makeatletter

\begin{document}

\title{Effects of counter-rotating-wave terms on the non-Markovianity in quantum open systems}

\author{Wei Wu}

\email{weiwu@csrc.ac.cn}

\affiliation{Beijing Computational Science Research Center, Beijing 100193, People's Republic of China}

\author{Maoxin Liu}

\affiliation{Beijing Computational Science Research Center, Beijing 100193, People's Republic of China}

\begin{abstract}
We investigate the effect of counter-rotating-wave terms on the non-Markovianity in quantum open systems by employing the hierarchical equations of motion in the framework of the non-Markovian quantum state diffusion approach. As illustrative examples, the non-Markovian memory effect of a qubit embedded in a bosonic or a fermionic environment with a detuned Lorentz spectrum at zero temperature is analyzed. It is found that the counter-rotating-wave terms are able to enhance the observed non-Markovianity whether the environment is composed of bosons or fermions. This result suggests that the rotating-wave approximation may reduce the non-Markovianity in quantum open systems. Moreover, we find that the modification of the non-Markovianity due to the different statistical properties of environmental modes becomes larger with the increase of the system-environment coupling strength.
\end{abstract}
\maketitle

\section{Introduction}\label{sec:sec1}

An accurate description of the dynamical behaviour of a quantum open system is one of the most challenging problems in the field of quantum mechanics~\cite{1}. The unavoidable coupling between the microcosmic quantum system and its surrounding environment is the main difficulty in solving this long-standing problem. The research of the dynamics of a quantum open system has attracted considerable attention in recent decades because it provides a possibility to simulate numerous physical and chemical processes. For example, the spin-boson model (SBM) and its extensions can be used to describe the excitation (electronic) energy transfer process in photosynthetic systems~\cite{2}. Traditionally, the dynamics of a quantum open system is treated by the perturbative theory along with the Markovian approximation~\cite{1,3}. This treatment is acceptable when the system-environment coupling is weak enough and the environment can be approximately described by a broadband spectrum. For the strong-coupling regime or the environment with memory, the Markovian approximation is no longer valid, and a more general non-Markovian dynamical formula is urgently required~\cite{3}.

When one speaks of a non-Markovian dynamical process, it implies that the dynamics is governed by a significant memory effect which has many applications in realistic physical systems~\cite{4,5}. In recent years, some rigorous theories or schemes for defining and measuring the non-Markovianity, i.e., the degree of non-Markovian memory effect in a dynamical process, have been proposed by different authors~\cite{4}. Many physical quantities are presented as the measure of the non-Markovianity in the previous literature, such as trace distance~\cite{5}, quantum Fisher information~\cite{6}, quantum mutual information~\cite{7}, quantum fidelity~\cite{8}, quantum channel capacity~\cite{9} and $k$-divisibility~\cite{10}. Using the trace distance to characterize the non-Markovian quantum behaviour is one of the most popular computable schemes. Trace distance is a metric of the distinguishability between two quantum states, the change in the trace distance can be interpreted as a flow of information between the quantum subsystem and the environment~\cite{5}. A Markovian process tends to continuously reduce the value of trace distance, or equivalently, the distinguishability between a pair of quantum states, which means that the information flows from the quantum subsystem to its environment. On the other hand, the increase of the distinguishability means a reversed flow of information from the environment back to the quantum subsystem which is the typical character of a non-Markovian quantum process.

Though the trace distance (as well as other measures) provides a computable characterization of the non-Markovianity for an arbitrary quantum open system, almost all the existing studies of the non-Markovianity have restricted their attention to some exactly solvable models~\cite{4,5,6,7,8,9,10}, say the pure dephasing model and the damped Jaynes-Cummings model~\cite{11}. This restriction is probably due to the fact that the calculation of the non-Markovianity requires an optimization over all the possible initial-state pairs. This  optimization procedure is rather time-consuming if the analytical expression of the quantum master equation describing the dynamical process is unknown. However, an analytical quantum master equation is typically obtained, in practice, via various approximations, e.g. the rotating-wave approximation (RWA) in which all the counter-rotating-wave terms are removed. On the other hand, it has been shown that the counter-rotating-wave terms play a very important role in the dynamics of the atomic population inversion (quantum Rabi oscillation)~\cite{12}, the quantum Zeno (and anti-Zeno) phenomenon for a hydrogen in free space~\cite{13}, the long-time evolution of the entanglement (quantum discord) in quantum open systems~\cite{14}, and the quantum phase transition between the Mott insulator and superfluid in a Rabi-Hubbard lattice~\cite{15}. An interesting question arises here: what are the influences of the counter-rotating-wave terms on the non-Markovianity in quantum open systems? In Ref.~\cite{16}, the authors tried to answer this question and showed that the RWA may lead to a dramatic reduction in the observed non-Markovianity in the SBM. Their conclusion is based on the second-order perturbative master equation technique which is valid only in the weak system-bath coupling regime. To exceed this limitation, a nonperturbative approach to the reduced dynamics of quantum open systems in the strong-coupling regime is requisite. In this paper, we employe the non-Markovian quantum state diffusion method along with a hierarchical equations of motion (HEOM)~\cite{17}, which is a highly-efficient and nonperturbative numerical approach, to reexamine their conclusion.

The non-Markovian stochastic Schr$\ddot{\mathrm{o}}$dinger equation of the quantum state diffusion type was developed by Strunz and his coworkers~\cite{18,19,20,21,22} which has shown momentous potential for solving dynamical problems in quantum open systems including both the bosonic~\cite{18,19,20}and the fermionic~\cite{21,22}environment situations. An obstacle of the non-Markovian quantum state diffusion method is that it contains not only the environmental noises, but also the functional derivative under a memory integral. This defect means that a directly numerical simulation of the non-Markovian quantum state diffusion equation needs a lot of CPU time. Very recently, it was shown that the numerical efficiency of the non-Markovian quantum state diffusion can be greatly improved by deriving a related hierarchy equation of reduced density matrices~\cite{17}. This result indicates that the non-Markovian quantum state diffusion approach is closely associated with the popular HEOM method which was widely used in many previous articles~\cite{23,24,25,26}, in other words, these two important approaches are completely equivalent under specified conditions. It is necessary to point out that this numerical treatment adopted in this paper (HEOM in the framework of the non-Markovian quantum state diffusion approach) includes all the orders of the system-bath interaction and is beyond the usual Markovian approximation, the RWA, and the perturbative approximation.

Most researchers are interested in how a quantum open system consisting of a small number of degrees of freedom, say a few-level system, interacts with an environment whose number of degrees of freedom tends to infinity. The decoherence or the relaxation of the quantum subsystem of interest is determined by the property of the environment to a great extent. Generally speaking, there are two kinds of quantum environments, one is the bosonic environment which is usually modeled by a set of harmonic oscillators, the other is the fermionic environment~\cite{addref1,addref2}, which can be realized by a spin chain via the well-known Jordan-Wigner transformation~\cite{addref2,27}. Most of the previous studies of the non-Markovianity only focused on the bosonic environment case. Do the bosonic environment and the fermionic environment have the same effect on the non-Markovianity? To address this question, we also extend our analysis to the fermionic environment situation. Thus, the other purpose of the present study is to compare the performances of these two kinds of environments in the non-Markovian behaviours of quantum open systems. It is also necessary to emphasize that such a spin-fermion model (SFM) is not merely of academic interest, but it connects with certain real physical matters, for example, the SFM may be used to describe the Ising-Kondo lattice with transverse magnetic field which is a possible candidate for the weak-moment heavy-fermion compound $\mathrm{UR}_{2}\mathrm{Si}_{2}$~\cite{28}.

This paper is organized as follows. In Sec.~\ref{sec:sec2}, we show how to derive the HEOM in the framework of the non-Markovian quantum state diffusion method. Moreover, we also briefly outline the general formalism of qualifying the non-Markovianity by using trace distance in a general quantum open system. In Sec.~\ref{sec:sec3}, we study the the effect of counter-rotating-wave terms on the non-Markovianity in both the SBM and the SFM. Some concerned discussions and the main conclusions of this paper are drawn in Sec.~\ref{sec:sec4}. Additionally, in the Appendix~\ref{sec:secapp}, we derive the exact quantum master equations of the SBM and the SFM under the RWA by using the non-Markovian quantum state diffusion method.

\section{Formulation}\label{sec:sec2}

In this section, we, first, would like to show how to derive the HEOM for the SBM and the SFM. The HEOM is a set of time-local differential equations for reduced density matrices of the quantum subsystem, which was originally proposed by Tanimura and his co-workers~\cite{23}. Secondly, we briefly outline the basic idea of the non-Markovianity measure in terms of the trace distance. For a better numerical performance, we made some slight modifications TO the definition of the non-Markovianity compared with the original one in Ref.~\cite{5}.

\subsection{The HEOM approach}\label{sec:sec2a}

Let us consider a quantum open system whose Hamiltonian can be described by
\begin{equation}\label{Eq:Eq1}
\hat{H}=\hat{H}_{s}+\sum_{k}\omega_{k}\hat{c}_{k}^{\dag}\hat{c}_{k}+\sum_{k}(g_{k}^{*}\hat{L}\hat{c}_{k}^{\dag}+g_{k}\hat{L}^{\dagger}\hat{c}_{k}),
\end{equation}
where $\hat{H}_{s}$ is the quantum subsystem of interest. In this paper, we assume $\hat{H}_{s}=\frac{1}{2}\epsilon\hat{\sigma}_{z}$ where $\epsilon$ is the transition frequency of the qubit system. Operator $\hat{L}$ denotes the quantum subsystem's operator coupled to the surrounding environment, and parameters $g_{k}$ are complex numbers quantifying the coupling strength between the quantum subsystem and its environment. Operators $\hat{c}_{k}$ and $\hat{c}_{k}^{\dagger}$ are the annihilation and creation operators of the $k$th environmental mode with frequency $\omega_{k}$, respectively. If $\hat{c}_{k}$ and $\hat{c}_{k}^{\dagger}$ satisfy the canonical commutation relations, i.e., $[\hat{c}_{k},\hat{c}_{k'}^{\dagger}]=\delta_{kk'}$, the Hamiltonian of Eq.~\ref{Eq:Eq1} refers to a extended SBM. On the other hand, if $\hat{c}_{k}$ and $\hat{c}_{k}^{\dagger}$ obey the canonical anti-commutation relations, i.e., $\{\hat{c}_{k},\hat{c}_{k'}^{\dagger}\}=\delta_{kk'}$, the whole system is a extended SFM.

The dynamics of the Hamiltonian $\hat{H}$ is governed by the Schr$\ddot{\mathrm{o}}$dinger equation $\partial_{t}|\Psi_{sb}(t)\rangle=-i\hat{H}|\Psi_{sb}(t)\rangle$, where $|\Psi_{sb}(t)\rangle$ is the wave function of the whole quantum open system. Practically, due to the large number of degrees of freedom in the environment, it is impossible to exactly solve this Schr$\ddot{\mathrm{o}}$dinger equation. However, by introducing the bosonic~\cite{29}or the fermionic~\cite{30}coherent-state $|\textbf{z}\rangle\equiv\bigotimes_{k}|z_{k}\rangle$ with $\hat{c}_{k}|z_{k}\rangle=z_{k}|z_{k}\rangle$, one can recast the Schr$\ddot{\mathrm{o}}$dinger equation into a stochastic Schr$\ddot{\mathrm{o}}$dinger equation of the quantum state diffusion type as follows~\cite{18,19,20,21,22}
\begin{equation}\label{Eq:Eq2}
\begin{split}
\frac{\partial}{\partial t}|\psi_{t}(\textbf{z}^{*})\rangle=&-i\hat{H}_{s}|\psi_{t}(\textbf{z}^{*})\rangle+\hat{L}\textbf{z}_{t}^{*}|\psi_{t}(\textbf{z}^{*})\rangle\\
&-\hat{L}^{\dagger}\int_{0}^{t}dsC(t-s)\frac{\delta}{\delta \textbf{z}_{s}^{*}}|\psi_{t}(\textbf{z}^{*})\rangle,
\end{split}
\end{equation}
where $|\psi_{t}(\textbf{z}^{*})\rangle\equiv\langle \textbf{z}|\Psi_{sb}(t)\rangle$ is the total pure-state wave function under the bosonic or the fermionic coherent-state representation, the variable $\textbf{z}_{t}\equiv i\sum_{k}g_{k}e^{-i\omega_{k}t}z_{k}$ can be interpreted as a stochastic process and satisfies $\mathcal{M}\{\textbf{z}_{t}\}=\mathcal{M}\{\textbf{z}_{t}^{*}\}=0$ and $\mathcal{M}\{\textbf{z}_{t}\textbf{z}_{s}^{*}\}=C(t-s)$ where $\mathcal{M}\{...\}$ denotes the statistical mean over all the possible stochastic processes, and $C(t)\equiv\sum_{k}|g_{k}|^{2}e^{-i\omega_{k}t}$ is the bath correlation function at zero temperature. It is convenient to encode the frequency dependence of the interaction strengths in the bath spectral function $J(\omega)\equiv\sum_{k}|g_{k}|^{2}\delta(\omega-\omega_{k})$, then the bath correlation function becomes $C(t)=\int d\omega J(\omega)e^{-i\omega t}$. The non-Markovian quantum state diffusion equation given by Eq.~\ref{Eq:Eq2} is applicable to an arbitrary bath spectral function $J(\omega)$, and no approximation is used in the derivation of  Eq.~\ref{Eq:Eq2}.

The explicit expression of the coherent-state $|z_{k}\rangle$ and the definition of the statistical mean operator $\mathcal{M}\{...\}$ have slight differences between the bosonic environment and the fermionic environment, however, these differences would not influence the derivation of Eq.~\ref{Eq:Eq2}. The physics behind the quantum state diffusion equation of Eq.~\ref{Eq:Eq2} is that the environment's effect on the quantum subsystem can be represented as stochastic noises exerting different influences on the quantum subsystem. These noises can have memory effects and are said to be colored which means these stochastic processes are non-Markovian. The main distinction between the SBM and the SFM is the mathematical properties of the noise entering the stochastic Schr$\ddot{\mathrm{o}}$dinger equation: the influence of a bosonic environment can be described exactly by a complex-valued colored Gaussian process~\cite{18,19,20}, while, a fermionic environment requires the use of a colored Grassmannian noise~\cite{21,22}. In the Markovian limit, the timescales of the environment are taken to be much shorter than the timescales of the quantum subsystem, all these stochastic noises reduce to white noises and the bath correlation function $C(t)$ becomes a Dirac-$\delta$ function which leads to a memoryless process~\cite{18,19,20,21,22}.

As shown in Eq.~\ref{Eq:Eq2}, the general non-Markovian quantum state diffusion equation contains a functional derivative with respect to the stochastic noise under a memory integral, which is not very convenient for a numerical simulation. However, if the bath correlation function can be expressed as a sum of exponential functions, this problem can be solved by making use of the idea of the HEOM~\cite{17,31}. In this paper, we consider the simplest case where the bath correlation function can be written as one exponential function, i.e.,
\begin{equation}\label{Eq:Eq4}
C(t)=\alpha e^{-\beta t},
\end{equation}
where $\alpha$ and $\beta$ are assumed to be complex numbers. For the more general case, i.e., $C(t)=\sum_{j}\alpha_{j} e^{-\beta_{j} t}$, this scheme still works~\cite{17,31}. Under the assumption in Eq.~\ref{Eq:Eq4}, one can replace the non-Markovian quantum state diffusion equation in Eq.~\ref{Eq:Eq2} by a set of hierarchial equations of the pure-state wave function $|\psi_{t}(\textbf{z}^{*})\rangle$. For the bosonic environment case, the hierarchical equations are given by~\cite{31}
\begin{equation}\label{Eq:Eq5}
\begin{split}
\frac{\partial}{\partial t}|\psi_{t}^{(m)}\rangle=&(-i\hat{H}_{s}-m\beta+\hat{L}\textbf{z}_{t}^{*})|\psi_{t}^{(m)}\rangle\\
&+m\alpha\hat{L}|\psi_{t}^{(m-1)}\rangle-\hat{L}^{\dagger}|\psi_{t}^{(m+1)}\rangle,
\end{split}
\end{equation}
where
\begin{equation*}
|\psi_{t}^{(m)}\rangle=|\psi_{t}^{(m)}(\textbf{z}^{*})\rangle\equiv\Bigg{[}\int_{0}^{t}dsC(t-s)\frac{\delta}{\delta \textbf{z}_{s}^{*}}\Bigg{]}^{m}|\psi_{t}(\textbf{z}^{*})\rangle,
\end{equation*}
are auxiliary pure-state wave functions. The hierarchy equation of pure-state wave functions in Eq.~\ref{Eq:Eq5} no longer contains the functional derivative, however, Eq.~\ref{Eq:Eq5} still contains the stochastic noise, which hinders the efficiency of a numerical simulation. To remove these stochastic noises, one needs to take the statistical mean over all the possible stochastic processes which is equivalent to tracing out the degree of the freedom of the environment, then the reduced density matrix of the quantum subsystem in the bosonic bath case is given by~\cite{18,19,20} $\hat{\varrho}_{s}(t)=\hat{\varrho}_{t}=\mathcal{M}\big{\{}|\psi_{t}(\textbf{z}^{*})\rangle\langle\psi_{t}(\textbf{z}^{*})|\big{\}}$. By making use of Eq.~\ref{Eq:Eq5} and the definition of $\hat{\varrho}_{s}(t)$, the HEOM of the SBM can be obtained as follows~\cite{17}
\begin{equation}\label{Eq:Eq7}
\begin{split}
\frac{d}{dt}\hat{\varrho}_{t}^{(m,n)}=&(-i\hat{H}_{s}^{\times}-m\beta-n\beta^{*})\hat{\varrho}_{t}^{(m,n)}\\
&+m\alpha\hat{L}\hat{\varrho}_{t}^{(m-1,n)}+n\alpha^{*}\hat{\varrho}_{t}^{(m,n-1)}\hat{L}^{\dag}\\
&-\hat{L}^{\dagger\times}\hat{\varrho}_{t}^{(m+1,n)}+\hat{L}^{\times}\hat{\varrho}_{t}^{(m,n+1)},
\end{split}
\end{equation}
where $\hat{X}^{\times}\hat{Y}\equiv [\hat{X},\hat{Y}]=\hat{X}\hat{Y}-\hat{Y}\hat{X}$ and $\hat{\varrho}_{t}^{(m,n)}\equiv\mathcal{M}\big{\{}|\psi_{t}^{(m)}(\textbf{z}^{*})\rangle\langle\psi_{t}^{(n)}(\textbf{z}^{*})|\big{\}}$ are auxiliary reduced density matrices. If $\hat{L}$ is a self-adjoint dissipation operator, i.e., $\hat{L}=\hat{L}^{\dagger}$, Eq.~\ref{Eq:Eq7} recovers the same HEOM which is derived by employing other methods, such as the stochastic decoupling scheme in Ref.~\cite{25} and the Feynman-Vernon influence functional approach in Ref.~\cite{23}.

Similarly, the hierarchy equation of pure-state wave functions for the SFM is given by~\cite{17}
\begin{equation}\label{Eq:Eq8}
\begin{split}
\partial_{t}|\psi_{t}^{(m)}\rangle=&[-i\hat{H}_{s}-m\beta+(-1)^{m}\hat{L}\textbf{z}_{t}^{*}]|\psi_{t}^{(m)}\rangle\\
&+\Theta(m)\alpha \hat{L}|\psi_{t}^{(m-1)}\rangle-\hat{L}^{\dagger}|\psi_{t}^{(m+1)}\rangle,
\end{split}
\end{equation}
where $\Theta(x)\equiv x~\mathrm{mod}~2$. Due to the anti-commutative multiplication induced by the Grassmannian process, it is very hard to directly numerically simulate Eq.~\ref{Eq:Eq8} which is quite different from that of the bosonic environment case. This problem can be naturally eliminated by taking the average over all the realizations of the noises, the reduced density matrix of the quantum subsystem in the fermionic environment case is~\cite{21,22} $\hat{\varrho}_{s}(t)=\hat{\varrho}_{t}=\mathcal{M}\big{\{}|\psi_{t}(\textbf{z}^{*})\rangle\langle\psi_{t}(-\textbf{z}^{*})|\big{\}}$, then, the HEOM of the SFM is given by~\cite{17}
\begin{equation}\label{Eq:Eq10}
\begin{split}
\frac{d}{dt}\hat{\varrho}^{(m,n)}_{t}&=(-i\hat{H}_{s}^{\times}-m\beta-n\beta^{*})\hat{\varrho}^{(m,n)}_{t}\\
&+\Theta(m)\alpha \hat{L}\hat{\varrho}^{(m-1,n)}_{t}+\Theta(n)\alpha^{*} \hat{\varrho}^{(m,n-1)}_{t}\hat{L}^{\dag}\\
&+[(-1)^{n}\hat{\varrho}^{(m+1,n)}_{t}\hat{L}^{\dagger}-\hat{L}^{\dagger}\hat{\varrho}^{(m+1,n)}_{t}]\\
&+[(-1)^{m}\hat{L}\hat{\varrho}^{(m,n+1)}_{t}-\hat{\varrho}^{(m,n+1)}_{t}\hat{L}],
\end{split}
\end{equation}
where auxiliary reduced density matrices are defined as $\hat{\varrho}_{t}^{(m,n)}=\mathcal{M}\big{\{}|\psi_{t}^{(m)}(\textbf{z}^{*})\rangle\langle\psi_{t}^{(n)}(-\textbf{z}^{*})|\big{\}}$.

The initial-state conditions of the auxiliary matrices for both the SBM and the SFM are $\hat{\varrho}^{(0,0)}_{t}=\hat{\varrho}_{s}(0)$ and $\hat{\varrho}^{(m>0,n>0)}_{t}=0$. In numerical simulations, we need to truncate the number of hierarchical equations for a sufficiently large integer $N$, which means all the terms of $\hat{\varrho}^{(m,n)}_{t}$ with $m+n>N$ are set to be zero. Then terms of $\hat{\varrho}^{(m,n)}_{t}$ with $m+n\leq N$ form a closed set of ordinary differential equations which can be solved directly by using the traditional Runge-Kutta method. It is necessary to point out that the HEOM approach is independent of the usual Markovian approximation, the RWA, and the perturbative approximation; in this sense, it can be regarded as a rigorous numerical method.

\subsection{The measure of non-Markovianity}\label{sec:sec2b}

In this paper, we adopt the trace distance as the quantity to characterize the non-Markovian memory effect in quantum open systems. One of the most important features of the non-Markovianity is the emergence of the recoherence or the information backflow from the environment to the subsystem which can be reflected by the rate of change in the trace distance between two physical initial states. The trace distance of two quantum states $\hat{\rho}_{1}$ and $\hat{\rho}_{2}$ is defined by~\cite{5} $D(\hat{\rho}_{1},\hat{\rho}_{2})\equiv\frac{1}{2}\|\hat{\rho}_{1}-\hat{\rho}_{2}\|_{1}$, where $\|\hat{X}\|_{1}\equiv\mathrm{Tr}\sqrt{\hat{X}^{\dagger}\hat{X}}$ is the trace norm or the Schatten one-norm of an arbitrary operator $\hat{X}$. For a initial-state pair $\hat{\rho}_{1,2}(0)$ and a given dynamical map $\hat{\Lambda}_{t}$ that generates the time-evolution $\hat{\rho}(t)=\hat{\Lambda}_{t}[\hat{\rho}(0)]$, one can define the rate of change of the trace distance as follows $\varpi[t;\hat{\rho}_{1,2}(0)]\equiv\frac{d}{dt}D[\hat{\rho}_{1}(t),\hat{\rho}_{2}(t)]$. When $\varpi[t;\hat{\rho}_{1,2}(0)]<0$, $\hat{\rho}_{1}(t)$ and $\hat{\rho}_{2}(t)$ approach each other, and this can be understood as the quantum information flows from the quantum subsystem to the environment; when $\varpi[t;\hat{\rho}_{1,2}(0)]>0$, $\hat{\rho}_{1}(t)$ and $\hat{\rho}_{2}(t)$  are away from each other, and this can be interpreted as the quantum information flows back to the quantum subsystem. In this spirit, a measure for the non-Markovianity of a quantum process can be defined by~\cite{5}
\begin{equation}\label{Eq:Eq13}
\mathcal{N}\equiv\max_{\hat{\rho}_{1,2}(0)}\int_{\varpi>0}dt\varpi[t;\hat{\rho}_{1,2}(0)],
\end{equation}
where the time-integration is extended over all time intervals $t\in[0,+\infty)$ in which $\varpi[t;\hat{\rho}_{1,2}(0)]$ is positive, and the maximum runs over all possible initial-state pairs $\hat{\rho}_{1,2}(0)$.

The definition of non-Markovianity in Eq.~\ref{Eq:Eq13} is not suitable for a numerical simulation in the following two aspects, first, it is impossible to numerically simulate the dynamics of the subsystem from zero to $+\infty$ which means we need a cutoff time $t_{c}$ for the time-integration in Eq.~\ref{Eq:Eq13}. Secondly, it is not very convenient to estimate whether or not $\varpi[t_{i};\hat{\rho}_{1,2}(0)]$ is positive or not at each given time $t=t_{i}$. Thus, we make some slight modifications in this paper: (i) we only focus on the non-Markovianity accumulated during a finite time interval $t\in[0,t_{c}]$, where $t_{c}$ is the upper bound of the time-integration; (ii) we change the original integrand and its corresponding integrating intervals in Eq.~\ref{Eq:Eq13} by a simple algebra that would not change the value of the non-Markovianity, then a equivalent expression of the non-Markovianity can be written as follows
\begin{equation}\label{Eq:Eq14}
\mathcal{N}\equiv\max_{\hat{\rho}_{1,2}(0)}\frac{1}{2}\int_{0}^{t_{c}}dt\big{\{}|\varpi[t;\hat{\rho}_{1,2}(0)]|+\varpi[t;\hat{\rho}_{1,2}(0)]\big{\}}.
\end{equation}
These modifications are also widely adopted in many previous studies~\cite{11,16}.

According to Refs.~\cite{16,32,33}, the calculation of the non-Markovianity $\mathcal{N}$ can be further simplified by choosing the two initial states as two orthogonal states that lie on the boundary of the space of physical states. For the qubit-system case, this orthogonality implies that both of the two initial states must be pure states~\cite{32}. Thus, in our numerical simulations, we assume the expressions of these initial-state pairs are given by $\hat{\rho}_{1}(0)=|\varphi_{s}(0)\rangle\langle\varphi_{s}(0)|$ and $\hat{\rho}_{2}(0)=|\varphi_{s}^{\perp}(0)\rangle\langle\varphi_{s}^{\perp}(0)|$ with
\begin{equation*}
|\varphi_{s}(0)\rangle=\cos\Big{(}\frac{\theta}{2}\Big{)}|e\rangle+e^{i\phi}\sin\Big{(}\frac{\theta}{2}\Big{)}|g\rangle,
\end{equation*}
and
\begin{equation*}
|\varphi_{s}^{\perp}(0)\rangle=\sin\Big{(}\frac{\theta}{2}\Big{)}|e\rangle-e^{i\phi}\cos\Big{(}\frac{\theta}{2}\Big{)}|g\rangle,
\end{equation*}
where $|e\rangle$ and $|g\rangle$ are the excited and the ground states of Pauli operator $\hat{\sigma}_{z}$, respectively, and the initial-state parameters $\theta\in[0,\pi]$ and $\phi\in[0,2\pi]$. By randomly generating a sufficiently large sample of initial-state parameter combinations $(\theta_{i},\phi_{i})$, one can find the optimal initial-state pair for the non-Markovianity. According to the definition in Eq.~\ref{Eq:Eq14}, the measure $\mathcal{N}$ is non-negative, and we have $\mathcal{N}=0$ if and only if the process is Markovian. A nonzero value $\mathcal{N}>0$, implies a non-Markovian process. It is noted that the non-Markovianity measure $\mathcal{N}$ represents a physically measurable quantity, and has been demonstrated in several recent experiments~\cite{34}.

\section{Results}\label{sec:sec3}

\begin{figure}
\centering
\includegraphics[angle=0,width=8cm]{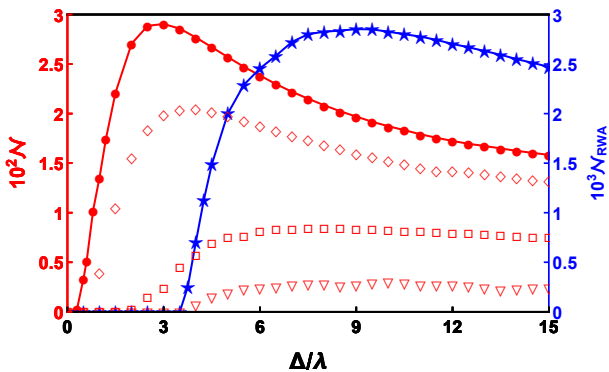}
\caption{\label{fig:fig1} The non-Markovianity $\mathcal{N}$ for the SBM with different initial-state parameters: optimal initial-state pair in our numerical simulation (filled red circles), $(\theta,\phi)=(\pi/8,0)$ (open red triangles), $(\theta,\phi)=(\pi/4,0)$ (open red squares) and $(\theta,\phi)=(3\pi/8,0)$ (open red diamonds). The filled blue stars are the result of the non-Markovianity under RWA with optimal initial-state parameters $(\theta,\phi)=(0,0)$. Other parameters are chosen as $\lambda=0.1$, $\gamma_{0}=0.02$, $\epsilon=2$, and the upper bound of the time-integration is $t_{c}=50$.}
\end{figure}

In this paper, we assume the dissipation operator of the quantum subsystem is given by $\hat{L}=\hat{\sigma}_{-}+\chi\hat{\sigma}_{+}$, where $\hat{\sigma}_{-}=\hat{\sigma}_{+}^{\dag}\equiv\frac{1}{2}(\hat{\sigma}_{x}-i\hat{\sigma}_{y})$. Parameter $\chi$ is a real number and $\chi\in[0,1]$. Then, the Hamiltonian given by Eq.~\ref{Eq:Eq1} can be recast as
\begin{equation}\label{Eq:Eq15}
\begin{split}
\hat{H}_{\chi}=&\frac{1}{2}\epsilon\hat{\sigma}_{z}+\sum_{k}\omega_{k}\hat{c}_{k}^{\dag}\hat{c}_{k}+\sum_{k}(g_{k}^{*}\hat{\sigma}_{-}\hat{c}_{k}^{\dag}+g_{k}\hat{\sigma}_{+}\hat{c}_{k})\\
&+\chi\sum_{k}(g_{k}^{*}\hat{\sigma}_{+}\hat{c}_{k}^{\dag}+g_{k}\hat{\sigma}_{-}\hat{c}_{k}),
\end{split}
\end{equation}
when $\chi=0$, all the counter-rotating-wave terms are removed and this Hamiltonian is under the RWA; while when $\chi=1$, all the contributions of the counter-rotating-wave terms are taken into consideration and this Hamiltonian is beyond the RWA. In this sense, the parameter $\chi$ stands for the strength of the counter-rotating-wave terms and can build a bridge between the RWA regime and the non-RWA regime. The introduction of the parameter $\chi$ helps us to get a deeper understanding of the effect of counter-rotating-wave terms on the non-Markovianity in quantum open systems.

In our numerical simulations, we assume the environment is initially prepared in its vacuum state $\bigotimes_{k}|0_{k}\rangle$ and the bath density spectral function $J(\omega)$ has the Lorentz spectrum form~\cite{18,19,20,21,22}
\begin{equation}\label{Eq:Eq16}
J_{\mathrm{L}}(\omega)=\frac{1}{2\pi}\frac{\gamma_{0}\lambda^{2}}{(\omega-\epsilon+\Delta)^{2}+\lambda^{2}},
\end{equation}
where $\lambda$ defines the spectral width of the coupling, $\gamma_{0}$ can be approximately interpreted as the system-bath coupling strength, and $\Delta$ is the detuning parameter which makes a shift from the transition frequency of the qubit system. Under this definition of the spectral function in Eq.~\ref{Eq:Eq16}, the bath correlation function $C(t)$ is given by
\begin{equation}\label{Eq:Eq17}
C_{\mathrm{L}}(t)=\frac{1}{2}\gamma_{0}\lambda \exp[-(\lambda+i\epsilon-i\Delta)t].
\end{equation}
It is quite obvious to see that the bath correlation function $C_{\mathrm{L}}$ satisfies the requirement (namely Eq.~\ref{Eq:Eq4}) to perform the HEOM scheme. In order to compare with previous studies reported in Refs.~\cite{5,16}, in this paper, we discuss the dependence of the non-Markovianity on the value of the detuning parameter $\Delta$.

\subsection{The bosonic environment case}\label{sec:sec3a}

\begin{figure}
\centering
\includegraphics[angle=0,width=8cm]{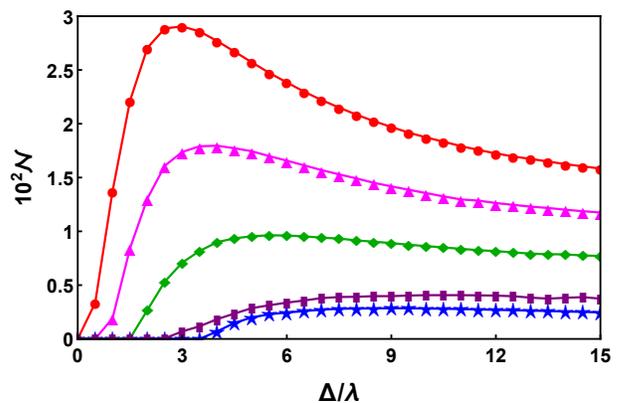}
\caption{\label{fig:fig2} The non-Markovianity $\mathcal{N}$ for the SBM with different $\chi$s: $\chi=1$ (red circles), $\chi=0.75$ (magenta triangles), $\chi=0.5$ (green diamonds), $\chi=0.25$ (purple squares) and $\chi=0$ (blue stars). Other parameters are chosen the same with Fig.~\ref{fig:fig1}.}
\end{figure}
\begin{figure}
\centering
\includegraphics[angle=0,width=8cm]{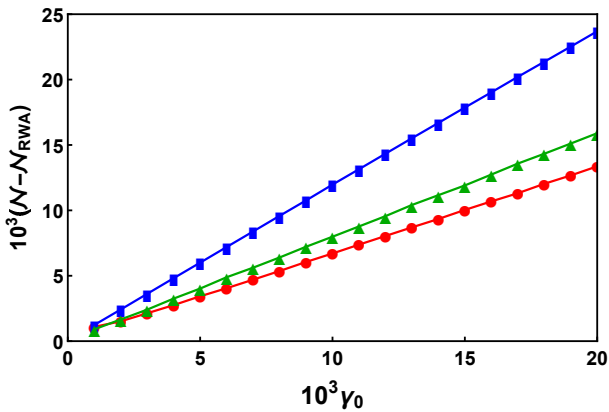}
\caption{\label{fig:fig3} $\mathcal{N}-\mathcal{N}_{\mathrm{RWA}}$ versus the system-bath coupling constant $\gamma_{0}$ for the SBM with different detuning parameters: $\Delta=5\lambda$ (blue squares), $\Delta=10\lambda$ (green triangles), $\Delta=15\lambda$ (red circles). Other parameters are chosen the same with Fig.~\ref{fig:fig1}.}
\end{figure}

In this subsection, we consider the bosonic environment case. For the RWA case $\chi=0$, our numerical simulations tell us that the maximization of the non-Markovianity $\mathcal{N}$ is attained for the initial state parameter $\theta=\phi=0$ which is consistent with previous studies~\cite{5}. As you can see in Fig.~\ref{fig:fig1}, the non-Markovianity $\mathcal{N}$ remains zero when $\Delta$ is very small and becomes nonzero at certain critical point $\Delta_{c}$ (in Fig.~\ref{fig:fig1}, $\Delta_{c}\simeq 3.5\lambda$), in which it first increases then decreases with the detuning parameter $\Delta$. This result indicates that the decoherence or the relaxation process in the RWA case is Markovian when $0<\Delta<\Delta_{c}$ and the non-Markovianity occurs for $\Delta>\Delta_{c}$. A crossover between Markovian and non-Markovian regimes appears at $\Delta=\Delta_{c}$.

For the non-RWA case $\chi=1$, we find that the value of the non-Markovianty $\mathcal{N}$ is obviously larger than that of the RWA case, especially in the regions where $\Delta$ is small. Thus, the non-RWA case displays a stronger non-Markovian phenomenon compared with that of the RWA situation. It is also shown that the decoherence or the relaxation process in the non-RWA case is non-Markovian for the entire range of $\Delta\in[0,15\lambda]$ (in Fig.~\ref{fig:fig1}, the non-Markovianity $\mathcal{N}$ at $\Delta=0$ is very close to zero, but is still positive). This phenomenon is very interesting, because it suggests that the SBM without the RWA generically exhibits non-Markovian behaviour. The same result is also noted in Ref.~\cite{35} in which the authors demonstrated that the ``eternal" non-Markovianity is typical for the SBM beyond the RWA. Strikingly, the non-RWA case yields at least one order of magnitude greater non-Markovianty compared with that obtained from the RWA. Moreover, we also consider the cases $\chi\in(0,1)$, as you can see in Fig.~\ref{fig:fig2}, with the increase of $\chi$ which means that with the enhancement of the influence of the counter-rotating-wave terms, the non-Markovianty $\mathcal{N}$ becomes larger. These results suggest that the counter-rotating-wave terms may enhance the non-Markovianty which is in agreement with previous results~\cite{16,35}.

It is well accepted that the RWA can be regarded as a good approximation if the system-environment coupling strength is very weak~\cite{1,36}, this result implies that the difference of the non-Markovianity induced by the counter-rotating-wave terms should vanish with the decease of the system-environment coupling strength. To check this conclusion, we plot the $\mathcal{N}-\mathcal{N}_{\mathrm{RWA}}$ as the function of the coupling constant $\gamma_{0}$ in Fig.~\ref{fig:fig3}. As you can see from the Fig.~\ref{fig:fig3}, the value of $\mathcal{N}-\mathcal{N}_{\mathrm{RWA}}$ linearly decreases as the system-bath coupling strength becomes small. This result meets our expectation and demonstrates that the influence of the counter-rotating-wave terms on the non-Markovianty is negligible only in the weak-coupling regime.

\subsection{The fermionic environment case}\label{sec:sec3b}

\begin{figure}
\centering
\includegraphics[angle=0,width=8cm]{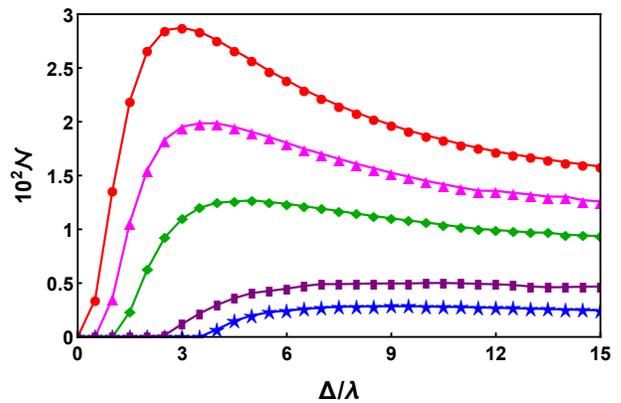}
\caption{\label{fig:fig4} The non-Markovianity $\mathcal{N}$ for the SFM with different $\chi$s: $\chi=1$ (red circles), $\chi=0.8$ (magenta triangles), $\chi=0.6$ (green diamonds), $\chi=0.3$ (purple squares) and $\chi=0$ (blue stars). Other parameters are chosen the same with Fig.~\ref{fig:fig1}.}
\end{figure}
\begin{figure}
\centering
\includegraphics[angle=0,width=8cm]{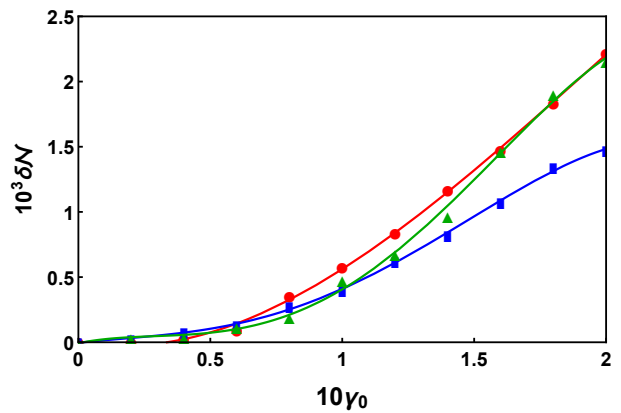}
\caption{\label{fig:fig5} $\delta\mathcal{N}\equiv|\mathcal{N}_{\mathrm{B}}-\mathcal{N}_{\mathrm{F}}|$ beyond the RWA versus the system-bath coupling constant $\gamma_{0}$ with different $\Delta$: $\Delta=5\lambda$ (green triangles), $\Delta=10\lambda$ (blue squares), $\Delta=15\lambda$ (red ciecles). Other parameters are chosen the same with Fig.~\ref{fig:fig1}.}
\end{figure}

Almost all the existing studies of the non-Markovianity in a quantum open system have restricted their attention to the SBM and to whether the quantum statistical property of environmental modes has effects on the non-Markovianity. To address this question, we extend our analysis to the fermionic environment situation in this subsection.

As you can see from Fig.~\ref{fig:fig4}, we find that the consideration of the counter-rotating-wave terms in the SFM may greatly enhance the non-Markovianity which is similar to the SBM case. Moreover, we also confirm that the difference of the non-Markovianity induced by the counter-rotating-wave terms linearly grows with the increase of the system-environment coupling strength in the fermionic environment case. Our numerical simulations show that the non-Markovianity $\mathcal{N}$ of the SFM under the RWA has the same value with that of the SBM case. In other words, the value of the non-Markovianity $\mathcal{N}$ is independent of the quantum statistical property of environmental modes under the RWA. In fact, one exactly derives the quantum master equations for the SBM and the SFM under the RWA. As we show in the Appendix~\ref{sec:secapp}, these two quantum master equations are the same at zero temperature which means the decoherence or the relaxation processes of the SBM and the SFM are identical in the RWA case. Similar results are also reported in Refs.~\cite{37} in which the authors found that the SBM (a two-level system coupled to a set of harmonic oscillators) and the SFM (a two-level system coupled to a set of spins which are equivalent to fermions with a Jordan-Wigner transformation) share the same dynamical behaviour at zero temperature.

Next we explore the influence of the quantum statistical property of environmental modes (Bose-Einstein statistics or Fermi-Dirac statistics) on the non-Markovianity beyond the RWA. We define a quantity $\delta\mathcal{N}\equiv|\mathcal{N}_{\mathrm{B}}-\mathcal{N}_{\mathrm{F}}|$ to detect the modification of the non-Markovianity due to the different quantum statistical properties of environmental modes. In Fig.~\ref{fig:fig5}, we display the $\delta\mathcal{N}$ as the function of the system-environment coupling strength $\gamma_{0}$. It is shown that the value of $\delta\mathcal{N}$ becomes large as the increase of $\gamma_{0}$, which implies that the modification of the non-Markovianity due to the different quantum statistical properties of environment modes maybe very obvious in the strong-coupling regime. Our finding also suggests that the reduced dynamics in a bosonic environment and a fermionic environment is equivalent in the weak-coupling regime. This result is in agreement with that of Ref.~\cite{38}.

This interesting phenomenon can be physically understood as follows: the main difference between the bosonic environment and the fermionic environment can be traced back to the Pauli exclusion principle. For the fermionic environment, the Pauli exclusion principle restricts the multi-fermion excitation process in each individual environmental mode, say $|0\rangle_{k}\rightarrow (c_{k}^{\dagger})^{\ell}|0\rangle_{k}$ is definitely forbidden if $\ell\geq 2$. While for the bosonic environment, such a multi-boson excitation process is allowed without the upper limit of $\ell$. Generally speaking, the multi-boson excitation process often happens in the strong-coupling regime or at high temperature, thus the dynamical behaviours of the SBM and the SFM can be regarded identical in the weak-coupling regime or at low temperature due to the fact that multi-boson excitation process is negligible in these two situations. This explanation is consistent with discussions shown in the appendix~\ref{sec:secapp}: the RWA in the SBM naturally forbids these multi-boson excitation processes if the environment is initially prepared in its vacuum state $\bigotimes_{k}|0_{k}\rangle$ (the RWA requires the total excitation number should be conservative), that is why, under the RWA, the SBM and the SFM share the same quantum master equation. However, at high temperature or without the RWA, such a restriction disappears, we expect that the dynamical behaviors of the SBM and the SFM are distinctly different, this conclusion is also in agreement with previous studies~\cite{37}.

\section{Discussions and Conclusions}\label{sec:sec4}

Here, we would like to have a brief discussion about the introduction of the parameter $\chi$ in Eq.~\ref{Eq:Eq15}. Most of the existing articles of quantum open systems only focus on two particular limits: the system totally with and without the RWA. In this paper, we let the strength of the counter-rotating-wave terms be tunable by introducing the parameter $\chi$ which is a very intuitive way to discuss the relationship between the counter-rotating-wave terms and the non-Markovianity. The parameter $\chi$ leads to an extended model in which we can bridge these two particular limits in a continuous path. Furthermore, we take the parameter $\chi$ into account for being not only motivated by theoretical curiosity but also inspired by the related discussion in the single mode version of SBM, namely the quantum Rabi model~\cite{add1}. In fact, for this light-matter coupled system, the tunable strength of the counter-rotating-wave terms leads to the so-called anisotropy~\cite{add2}. It has been found that the existence of anisotropy can significantly affect many interesting properties of the quantum Rabi model, e.g., the universality of the quantum phase transition~\cite{add3}, the quantum Fisher information~\cite{add4} and the squeezing of the light field~\cite{add5}. There are already several proposals to realize the anisotropic form coupling in some potential experimental candidates, such as the cavity quantum electrodynamics (QED), the superconducting circuit and the spin-orbit-coupling systems~\cite{add2}. In Ref.~\cite{add2}, the authors found that the experimental data of the Bloch-Siegert shift in the ultrastrong-coupling regime can be perfectly fitted in terms of the anisotropic Rabi model which is neither totally with nor without the RWA. There is even a proposal for the realization of the anisotropic counter-rotating-wave coupling in a many-body system~\cite{add6}. Therefore, we are motivated to explore the non-Markovian effect in quantum open systems in the way that the strength of the counter-rotating-wave terms is controlled by the tunable parameter $\chi$.

In conclusion, we investigate the effect of counter-rotating-wave terms on the non-Markovianity in quantum open systems by employing the non-Markovian quantum state diffusion approach along with the HEOM. This numerical technique is applicable for both the SBM and the SFM without the usual Markovian, rotating-wave and perturbative approximations. It is found that the counter-rotating-wave terms are able to enhance the observed non-Markovianity in quantum open systems no matter the environment is composed of bosons or fermions. This result suggests that the RWA may inherently or inevitably introduce the Markovian assumption. We also find that the modification of the non-Markovianity induced by the counter-rotating-wave terms, i.e., $\mathcal{N}-\mathcal{N}_{\mathrm{RWA}}$, becomes small as the system-bath coupling strength decreases regardless of the SBM or the SFM. This finding means the influence of the counter-rotating-wave terms can be ignored in the weak-coupling regime which is consistent with many previous studies about the application scope of the RWA. Moreover, it is shown that the deviation of the non-Markovianity due to the different quantum statistical properties of environmental modes, i.e., $\delta \mathcal{N}$, becomes large with the increase of system-bath coupling strength, which implies that the influence of quantum statistical properties of environmental modes on the non-Markovianity of quantum open systems becomes significant in the strong-coupling regime.

Though these results are achieved in the detuned Lorentz spectrum, our numerical scheme can be expected to generalize to other bath spectral functions. In recent years, the HEOM method has been extended to arbitrary spectral functional forms~\cite{39}, it would be very interesting to recheck the relationship between the non-Markovianity and the RWA in other bath spectral density functions $J(\omega)$ cases, such as the sub-Ohmic, Ohmic and super-Ohmic spectrums. In addition, the non-Markovian quantum state diffusion method can be generalized to the finite-temperature environment situation~\cite{19,40}, in which one can map the finite-temperature environment onto a larger zero-temperature environment by using the thermofield method~\cite{41} which doubles the number of the stochastic processes required. This result indicates that the numerical scheme proposed in this paper can be also adopted to explore the effect of environmental temperature on the non-Markovianity. Finally, due to the generality of the quantum open system, we expect our results to be of interest for a wide range of experimental applications in quantum computation and quantum information processing.

\section{Acknowledgments}\label{sec:secack}

W. W. and M. L. wish to thank Dr. Da-Wei Luo, Prof. Hai-Qing Lin, Prof. Jian-Qiang You and Prof. Ting Yu for many useful discussions. W. W. is supported by the China Postdoctoral Science Foundation (Grant No.2017M610753) and the NSFC (Grant No.11704025), M. L. is supported by the NSFC (Grant No.11604009), W. W. and M. L. acknowledge the support from NSAF U1530401 and the computational resource from the Beijing Computational Science Research Center.

\section{Appendix}\label{sec:secapp}

In this appendix, we would like to derive the exact master equation of Eq.~\ref{Eq:Eq15} under the RWA by employing the non-Markovian quantum state diffusion approach. Regardless of the bosonic or the fermioninc environment cases, the non-Markovian quantum state diffusion equation for the Hamiltonian in Eq.~\ref{Eq:Eq1} is given by Eq.~\ref{Eq:Eq2}. The main difficulty in solving this stochastic differential equation comes from the exact treatment of the functional derivation. Following the scheme proposed in Refs.~\cite{18,19,20,21,22}, one can replace the functional derivative with a time-dependent operator of the form
\begin{equation}\label{Eq:Eq18}
\frac{\delta}{\delta \textbf{z}_{s}^{*}}|\psi_{t}(\textbf{z}^{*})\rangle=\mathcal{\hat{O}}(t,s,\textbf{z}^{*})|\psi_{t}(\textbf{z}^{*})\rangle,
\end{equation}
and the $\mathcal{\hat{O}}(t,s,\textbf{z}^{*})$-operator can be functional expanded as follows
\begin{equation}\label{Eq:Eq19}
\begin{split}
\mathcal{\hat{O}}(t,s,&\textbf{z}^{*})=\mathcal{\hat{O}}_{0}(t,s)+\int_{0}^{t}\mathcal{\hat{O}}_{1}(t,s,\nu_{1})\textbf{z}_{\nu_{1}}^{*}d\nu_{1}\\
&+\int_{0}^{t}\int_{0}^{t}\mathcal{\hat{O}}_{2}(t,s,\nu_{1},\nu_{2})\textbf{z}_{\nu_{1}}^{*}\textbf{z}_{\nu_{2}}^{*}d\nu_{1}d\nu_{2}+...
\end{split}
\end{equation}
The necessary condition for the $\mathcal{\hat{O}}(t,s,\textbf{z}^{*})$-operator to contain a finite number of noise-dependent terms in this expansion is $\hat{L}^{\times}\hat{L}^{\times}...\hat{L}^{\times}\hat{H}_{s}=0$. And the equation of motion of $\mathcal{\hat{O}}_{n}(t,s,\nu_{1},\nu_{2},...,\nu_{n})$ are determined by the following consistency condition
\begin{equation}\label{Eq:Eq20}
\frac{\partial}{\partial t}\frac{\delta}{\delta \textbf{z}_{s}^{*}}|\psi_{t}(\textbf{z}^{*})\rangle=\frac{\delta}{\delta \textbf{z}_{s}^{*}}\frac{\partial}{\partial t}|\psi_{t}(\textbf{z}^{*})\rangle,
\end{equation}
where initial conditions are given by $\mathcal{\hat{O}}_{0}(t,t)=\hat{L}$ and $\mathcal{\hat{O}}_{n}(t,t,\nu_{1},\nu_{2},...,\nu_{n})=0$ for $n\geq 1$.

For the RWA case, i.e., $\hat{H}_{s}=\frac{1}{2}\epsilon\hat{\sigma}_{z}$ and $\hat{L}=\hat{\sigma}_{-}$, only the zero-order terms in the functional expansion are required~\cite{18,19,20,21,22}, namely, $\mathcal{\hat{O}}(t,s,\textbf{z}^{*})=\mathcal{\hat{O}}_{0}(t,s)=f(t,s)\hat{\sigma}_{-}$, where $f(t,s)$ is a unknown time-dependent coefficient. By substituting the above expression into the consistency condition, one can find that $f(t,s)$ needs to satisfy the following self-consistent equation:
\begin{equation}\label{Eq:Eq22}
\frac{\partial}{\partial t}f(t,s)=[i\epsilon+F(t)]f(t,s),
\end{equation}
where
\begin{equation*}
F(t)=\int_{0}^{t}C(t-s)f(t,s)ds,
\end{equation*}
with $f(t,s=t)=1$. Then, the non-Markovian quantum state diffusion equation in Eq.~\ref{Eq:Eq2} can be rewritten by
\begin{equation*}
\begin{split}
\frac{d}{d t}|\psi_{t}(\textbf{z}^{*})\rangle=&-\frac{i}{2}\epsilon\hat{\sigma}_{z}|\psi_{t}(\textbf{z}^{*})\rangle+\hat{\sigma}_{-}\textbf{z}_{t}^{*}|\psi_{t}(\textbf{z}^{*})\rangle\\
&-F(t)\hat{\sigma}_{+}\hat{\sigma}_{-}|\psi_{t}(\textbf{z}^{*})\rangle,
\end{split}
\end{equation*}

The deterministic master equation can be derived from the above stochastic differential equation by statistical mean over all the possible stochastic processes. For the bosonic environment case,
\begin{equation*}
\begin{split}
\frac{d}{dt}\hat{\varrho}_{t}=&\mathcal{M}\Bigg{\{}\frac{\overrightarrow{d}}{dt}|\psi_{t}(\textbf{z}^{*})\rangle\langle\psi_{t}(\textbf{z}^{*})|\Bigg{\}}\\
&+\mathcal{M}\Bigg{\{}|\psi_{t}(\textbf{z}^{*})\rangle\langle\psi_{t}(\textbf{z}^{*})|\frac{\overleftarrow{d}}{dt}\Bigg{\}}\\
=&-\frac{i}{2}\epsilon\hat{\sigma}_{z}\hat{\varrho}_{t}+\frac{i}{2}\epsilon\hat{\varrho}_{t}\hat{\sigma}_{z}\\
&+\hat{\sigma}_{-}\mathcal{M}\big{\{}\textbf{z}_{t}^{*}|\psi_{t}(\textbf{z}^{*})\rangle\langle\psi_{t}(\textbf{z}^{*})|\big{\}}-F(t)\hat{\sigma}_{+}\hat{\sigma}_{-}\hat{\varrho}_{t}\\
&+\mathcal{M}\big{\{}|\psi_{t}(\textbf{z}^{*})\rangle\langle\psi_{t}(\textbf{z}^{*})|\textbf{z}_{t}\big{\}}\hat{\sigma}_{+}-F^{*}(t)\hat{\varrho}_{t}\hat{\sigma}_{+}\hat{\sigma}_{-},
\end{split}
\end{equation*}
where $\overleftarrow{d}$ and $\overrightarrow{d}$ are the left and right time derivative with respect to $|\psi_{t}(\textbf{z}^{*})\rangle$, respectively. The above equation can be further simplified by making use of the Novikov's theorem~\cite{20}
\begin{equation*}\label{Eq:Eq23}
\begin{split}
\mathcal{M}&\big{\{}|\psi_{t}(\textbf{z}^{*})\rangle\langle\psi_{t}(\textbf{z}^{*})|\textbf{z}_{t}\big{\}}\\
&=\mathcal{M}\Bigg{\{}\int_{0}^{t}dsC(t-s)\frac{\overrightarrow{\delta}}{\delta\textbf{z}_{s}^{*}}|\psi_{t}(\textbf{z}^{*})\rangle\langle\psi_{t}(\textbf{z}^{*})|\Bigg{\}}\\
&=F(t)\hat{\sigma}_{-}\hat{\varrho}_{t},
\end{split}
\end{equation*}
and
\begin{equation*}\label{Eq:Eq24}
\begin{split}
\mathcal{M}&\big{\{}\textbf{z}_{t}^{*}|\psi_{t}(\textbf{z}^{*})\rangle\langle\psi_{t}(\textbf{z}^{*})|\big{\}}\\
&=\mathcal{M}\Bigg{\{}\int_{0}^{t}dsC^{*}(t-s)|\psi_{t}(\textbf{z}^{*})\rangle\langle\psi_{t}(\textbf{z}^{*})|\frac{\overleftarrow{\delta}}{\delta\textbf{z}_{s}^{*}}\Bigg{\}}\\
&=F^{*}(t)\hat{\varrho}_{t}\hat{\sigma}_{+},
\end{split}
\end{equation*}
where $\overleftarrow{\delta}$ and $\overrightarrow{\delta}$ are the left and right functional derivative with respect to $\textbf{z}_{s}^{*}$, respectively. Finally, the exact quantum master equation of the SBM with the RWA is given by
\begin{equation}\label{Eq:Eq25}
\begin{split}
\frac{d}{dt}\hat{\varrho}_{t}=&-\frac{i}{2}\epsilon[\hat{\sigma}_{z},\hat{\varrho}_{t}]+F(t)[\hat{\sigma}_{-}\hat{\varrho}_{t}\hat{\sigma}_{+}-\hat{\sigma}_{+}\hat{\sigma}_{-}\hat{\varrho}_{t}]\\
&+F^{*}(t)[\hat{\sigma}_{-}\hat{\varrho}_{t}\hat{\sigma}_{+}-\hat{\varrho}_{t}\hat{\sigma}_{+}\hat{\sigma}_{-}].
\end{split}
\end{equation}

If the spectral function $J(\omega)$ is the Lorentz spectrum with zero detuning, i.e., $\Delta=0$, the self-consistent equation of $F(t)$ in Eq.~\ref{Eq:Eq22} becomes
\begin{equation*}
\frac{d}{dt}F(t)=F^{2}(t)-\lambda F(t)+\frac{1}{2}\gamma_{0}\lambda,
\end{equation*}
the analytical expression of the above equation is
\begin{equation*}
\begin{split}
F(t)&=F^{*}(t)\\
&=\frac{1}{2}\Bigg{\{}\lambda-\Omega\tanh\Bigg{[}\frac{1}{2}\Omega t+ \mathrm{arc}\tanh\Bigg{(}\frac{\lambda}{\Omega}\Bigg{)}\Bigg{]}\Bigg{\}},
\end{split}
\end{equation*}
where $\Omega=\sqrt{\lambda^{2}-2\gamma_{0}\lambda}$, then the quantum master equation of Eq.~\ref{Eq:Eq25} recovers the result reported in Ref.~\cite{42} where the stochastic decoupling scheme proposed by Shao is adopted. The exact solution of Eq.~\ref{Eq:Eq25} can be expressed in the standard basis $\{|e\rangle,|g\rangle\}$ as follows
\begin{equation*}
\hat{\varrho}_{t}=\left[
                    \begin{array}{cc}
                      \varrho_{ee}(0)G^{2}(t) & \varrho_{eg}(0)G(t)e^{-i\epsilon t} \\
                      \varrho_{ge}(0)G(t)e^{i\epsilon t} & \varrho_{gg}(0)G^{2}(t) \\
                    \end{array}
                  \right],
\end{equation*}
where the decay factor is given by
\begin{equation*}
\begin{split}
G(t)=&-\int_{0}^{t}F(s)ds\\
=&\exp\Big{(}-\frac{1}{2}\lambda t\Big{)}\Big{[}\cosh\Big{(}\frac{1}{2}\Omega t\Big{)}+\frac{\lambda}{\Omega}\sinh\Big{(}\frac{1}{2}\Omega t\Big{)}\Big{]}.
\end{split}
\end{equation*}
Compared with previous results reported in Refs.~\cite{1,11}, a phase shift $e^{\pm i\epsilon t}$ occurs in the non-diagonal elements in our expression. However, considering the fact that the definition of the non-Markovianity adopted in this paper is invariant under a time-local unitary transformation~\cite{16}, one can easily eliminate this phase shift, which does not change the value of the non-Markovianity, by using $\hat{\varrho}_{t}\rightarrow \hat{U}(t)\hat{\varrho}_{t}\hat{U}^{\dag}(t)$ with $\hat{U}(t)=\exp(\frac{i}{2}\epsilon t\hat{\sigma}_{z})$. Then one can recover the same results in Refs.~\cite{1,11}, which convinces us that the non-Markovian quantum state diffusion approach truly captures the dynamical behaviour of a quantum open system.

For the fermionic environment case, the deterministic quantum master equation can be derived by
\begin{equation*}
\begin{split}
\frac{d}{dt}\hat{\varrho}_{t}=&\mathcal{M}\Bigg{\{}\frac{\overrightarrow{d}}{dt}|\psi_{t}(\textbf{z}^{*})\rangle\langle\psi_{t}(-\textbf{z}^{*})|\Bigg{\}}\\
&+\mathcal{M}\Bigg{\{}|\psi_{t}(\textbf{z}^{*})\rangle\langle\psi_{t}(-\textbf{z}^{*})|\frac{\overleftarrow{d}}{dt}\Bigg{\}}\\
=&-\frac{i}{2}\epsilon\hat{\sigma}_{z}\hat{\varrho}_{t}+\frac{i}{2}\epsilon\hat{\varrho}_{t}\hat{\sigma}_{z}\\
&+\hat{\sigma}_{-}\mathcal{M}\big{\{}\textbf{z}_{t}^{*}|\psi_{t}(\textbf{z}^{*})\rangle\langle\psi_{t}(-\textbf{z}^{*})|\big{\}}-F^{*}(t)\hat{\varrho}_{t}\hat{\sigma}_{+}\hat{\sigma}_{-}\\
&-\mathcal{M}\big{\{}|\psi_{t}(\textbf{z}^{*})\rangle\langle\psi_{t}(-\textbf{z}^{*})|\textbf{z}_{t}\big{\}}\hat{\sigma}_{+}-F(t)\hat{\sigma}_{+}\hat{\sigma}_{-}\hat{\varrho}_{t}.
\end{split}
\end{equation*}
Making use of the Novikov theorem of the Grassmannian noise~\cite{21,22}
\begin{equation*}\label{Eq:Eq26}
\begin{split}
\mathcal{M}&\big{\{}|\psi_{t}(\textbf{z}^{*})\rangle\langle\psi_{t}(-\textbf{z}^{*})|\textbf{z}_{t}\big{\}}\\
&=-\mathcal{M}\Bigg{\{}\int_{0}^{t}dsC(t-s)\frac{\overrightarrow{\delta}}{\delta\textbf{z}_{s}^{*}}|\psi_{t}(\textbf{z}^{*})\rangle\langle\psi_{t}(-\textbf{z}^{*})|\Bigg{\}}\\
&=-F(t)\hat{\sigma}_{-}\hat{\varrho}_{t},
\end{split}
\end{equation*}
and
\begin{equation*}\label{Eq:Eq27}
\begin{split}
&\mathcal{M}\big{\{}\textbf{z}_{t}^{*}|\psi_{t}(\textbf{z}^{*})\rangle\langle\psi_{t}(-\textbf{z}^{*})|\big{\}}\\
&=-\mathcal{M}\Bigg{\{}\int_{0}^{t}dsC^{*}(t-s)|\psi_{t}(\textbf{z}^{*})\rangle\langle\psi_{t}(-\textbf{z}^{*})|\frac{\overleftarrow{\delta}}{\delta\textbf{z}_{s}^{*}}\Bigg{\}}\\
&=\mathcal{M}\Bigg{\{}\int_{0}^{t}dsC^{*}(t-s)|\psi_{t}(\textbf{z}^{*})\rangle\langle\psi_{t}(-\textbf{z}^{*})|\frac{\overleftarrow{\delta}}{\delta(-\textbf{z}_{s}^{*})}\Bigg{\}}\\
&=F^{*}(t)\hat{\varrho}_{t}\hat{\sigma}_{+},
\end{split}
\end{equation*}
one can obtain that the exact quantum master equation of the SFM which has the same expression with that of Eq.~\ref{Eq:Eq25}. The same dynamical behaviour observed in the RWA case indicates that it is hard to identify a distinction between the bosonic and fermionic environments from the viewpoint of the reduced dynamics. The fact that the SBM and the SFM under the RWA share the same quantum master equation is not a coincidence, because the RWA restricts the multi-boson excitation process, i.e., $|0\rangle_{k}\rightarrow (c_{k}^{\dagger})^{\ell}|0\rangle_{k}$ with $\ell\geq 2$, at zero temperature which leads to the same structures of the bosonic and the fermionic environments. This result suggests that the RWA may eliminate the peculiarity induced by the quantum statistical property of environmental modes at zero temperature.

It is necessary to point out that the decoherence or the relaxation dynamics in bosonic and fermionic environments are generally different for the case $\chi\neq 0$. This is because the higher-order terms of the functional expansion in Eq.~\ref{Eq:Eq19} are also contribute to the dynamical behaviour. We strongly suspect that these higher-order terms are able to enhance the non-Markovianity regardless of the SBM or the SFM. In the non-RWA case, no exact closed master equation can be obtained by the process outlined in this appendix (the non-Markovian quantum state diffusion method) due the fact that the number of the functional expansion of the $\mathcal{\hat{O}}(t,s,\textbf{z}^{*})$-operator in Eq.~\ref{Eq:Eq19} is infinite (because $\hat{L}^{\times}\hat{L}^{\times}...\hat{L}^{\times}\hat{H}_{s}\neq0$ in the non-RWA case). However, it has been reported that an exact closed quantum master equation of the SBM without the RWA can be derived by employing the dynamical map technique proposed in Ref.~\cite{43}.

\end{document}